\documentclass{jfm}
\usepackage{graphicx}
\usepackage{epstopdf,epsfig}
\usepackage{newtxtext}
\usepackage{newtxmath}
\usepackage{natbib}
\usepackage{hyperref}
\hypersetup{
    colorlinks = true,
    urlcolor   = blue,
    citecolor  = black,
}

\newcommand{\RomanNumeralCaps}[1]

\usepackage{tikz}

\DeclareRobustCommand\dotted{\tikz[baseline=-0.6ex] \draw [thick, dotted, line width=0.5mm, black ] (0,0)--(0.54,0) ;}
\DeclareRobustCommand\dashed{\tikz[baseline=-0.6ex]\draw[thick,dashed] (0,0)--(0.54,0);}
\DeclareRobustCommand\chain {\tikz[baseline=-0.6ex]\draw[thick, dash dot dot, line width=0.5mm, black] (0,0)--(0.54,0);}
\AtBeginDocument{}
\usepackage{xcolor}
\definecolor{MATblue}{rgb}{0 0.4470 0.7410}
\definecolor{MATred}{rgb}{0.8500 0.3250 0.0980}
\usepackage{cancel}
\usepackage{mathtools}
\usepackage{amsmath}
\usepackage[utf8]{inputenc}
\usepackage{changes}
\usepackage{subcaption}

\let\realverbatim=\verbatim
\let\realendverbatim=\endverbatim
\renewcommand\verbatim{\par\addvspace{6pt plus 2pt minus 1pt}\realverbatim}
\renewcommand\endverbatim{\realendverbatim\addvspace{6pt plus 2pt minus 1pt}}

\title{A solution for the quasi-one-dimensional linearised Euler equations with heat transfer}

\author{Saikumar R. Yeddula, Juan Guzmán-Iñigo, Aimee S. Morgans\\
{\small \textit{Department of Mechanical Engineering, Imperial College London, London\\
e-mail: s.yeddula18@imperial.ac.uk}}}
\author{Saikumar R. Yeddula\aff{1}
  \corresp{\email{s.yeddula18@imperial.ac.uk}},
  Juan Guzmán-Iñigo\aff{1}
 \and Aimee S. Morgans\aff{1}}

\affiliation{\aff{1}STM Department of Mechanical Engineering, Imperial College London, London SW7 2AZ, UK}

\begin{document}
\maketitle

\begin{abstract}
The unsteady response of nozzles with steady heat transfer forced by acoustic and/or entropy waves is modelled. The approach is based on the quasi-one-dimensional linearised Euler equations. The equations are cast in terms of three variables, namely the dimensionless mass, stagnation temperature and entropy fluctuations, which are invariants of the system at zero frequency and with no heat transfer. The resulting first-order system of differential equations is then solved using the Magnus expansion method, where the perturbation parameters are the normalised frequency and the volumetric heat transfer. In this work, a measure of the flow non-isentropicity (in this case the steady heat transfer) is used for the first time as an expansion parameter. The solution method was applied to a converging–diverging nozzle with constant heat transfer for both sub-critical and super-critical flow cases, showing good agreement with numerical predictions. It was observed that the acoustic and entropy transfer functions of the nozzle strongly depend on the frequency and heat transfer.

\end{abstract}
\section{Introduction}
\label{sec:intro}
Solutions of the unsteady response of ducts with area variations and sustaining a mean flow, i.e. nozzle flows, are of interest for a wide variety of industrial applications, including combustors, automotive exhausts, after-burners, or supersonic air-intake diffusers. An early attempt to mathematically describe the transfer function of rocket nozzles was proposed by~\citet{tsien1952transfer}, motivated by emerging theoretical descriptions of combustion instabilities in rocket engines. 

With the same motivation,~\citet{Marble_JSV_1977} derived an analytical solution for the transfer function of both sub-critical and super-critical nozzles excited by acoustic and/or entropy waves. This solution is valid in the compact (or zero-frequency) limit which assumes that the acoustic and entropy wavelengths significantly exceed the length scale over which the nozzle area change occurs. This compact solution was extended to non-zero frequencies by~\citet{Stow_JFM_2002} and~\citet{Goh_JSV_2011} using an asymptotic expansion of the linearised Euler equations in terms of frequency. Building on the same assumptions of~\citet{Marble_JSV_1977}, namely inviscid, isentropic and quasi-one-dimensional flow,~\citet{Duran_JFM_2013} proposed a solution valid at any frequency based on the Magnus expansion method \citep{Blanes_PR_2009}. This approach was later extended to account for incoming compositional inhomogeneities by~\citet{magri2017indirect}, to annular nozzles by~\citet{duran2015reflection} and to multi-stream nozzles by~\citet{younes2019indirect}. All the previous studies were developed in the context of indirect combustion noise~\citep{ihme2017combustion}, which is the noise generated when convective disturbances (entropy, vortical and compositional waves) are accelerated/decelerated. Indirect combustion noise is particularly relevant for high Mach number flows and low frequencies. 

In parallel to the theoretical efforts of the combustion noise community, solutions 
for the acoustic field in nozzle flows
were presented for low subsonic Mach numbers.~\citet{Dokumaci_JSV_1998a} proposed
a semi-analytical solution based on the WKB method which requires sufficiently high frequencies. Additionally, several analytical solutions of the acoustic field in nozzles with specific area profiles were obtained by~\citet{Eisenberg_JASA_1971}, \citet{Easwaran_JSV_1992} and \citet{Subrahmanyam_JSV_2001} 
using transformation of variables. 

An assumption of all the aforementioned studies is the isentropicity of the flow. However, many practical applications require solutions for non-isentropic mean flows. For instance,~\citet{de2021compositional} and \citet{yang2020entropy} recently showed that, in the diverging portion of realistic nozzles, the flow can separate creating recirculation regions and, hence, turbulence both leading to mean flow non-isentropicity and strong deviations from isentropic acoustic theories. Another source of non-isentropicity can be heat transfer, as happens in combustors and heat exchangers.~\citet{Yeddula_JSV_2021} showed that heat transfer can significantly affect the acoustics of nozzles. However, there is a lack of solutions to this problem, even in the compact limit. The only solution to date was proposed by~\citet{Yeddula_JSV_2020} based on the WKB method and assumes high frequencies and low subsonic flow Mach numbers.

In this paper, we propose a general solution for the acoustic field produced by 
acoustic and/or entropy waves in a nozzle whose mean flow is non-isentropic due to steady heat transfer. The solution is based on the Magnus expansion method of~\citet{Duran_JFM_2013} and is valid for both sub-critical and super-critical flow conditions, and for any frequency within the limit of the 
quasi-one-dimensional assumption. The Magnus expansion when applied to nozzle flows has always so far had frequency as the expansion parameter. In this work, we include a measure of the flow non-isentropicity as an expansion parameter for the first time, allowing us to develop models for non-isentropic nozzle flows which are valid at all frequencies and Mach numbers. The non-isentropicity considered is due to axially varying steady heat transfer; the method could be extended to include other forms of non-isentropicity in future work.

This article is organized as follows. In~\S\ref{sec:Analysis} the mathematical model is described and a solution is proposed based on the Magnus expansion method. The solution is applied to a converging-diverging nozzle in~\S\ref{sec:Results} and conclusions are drawn in \S\ref{sec:Conclusions}. 
\section{Analysis}\label{sec:Analysis}
\begin{figure}
  \centerline{\includegraphics[width=0.7\linewidth]{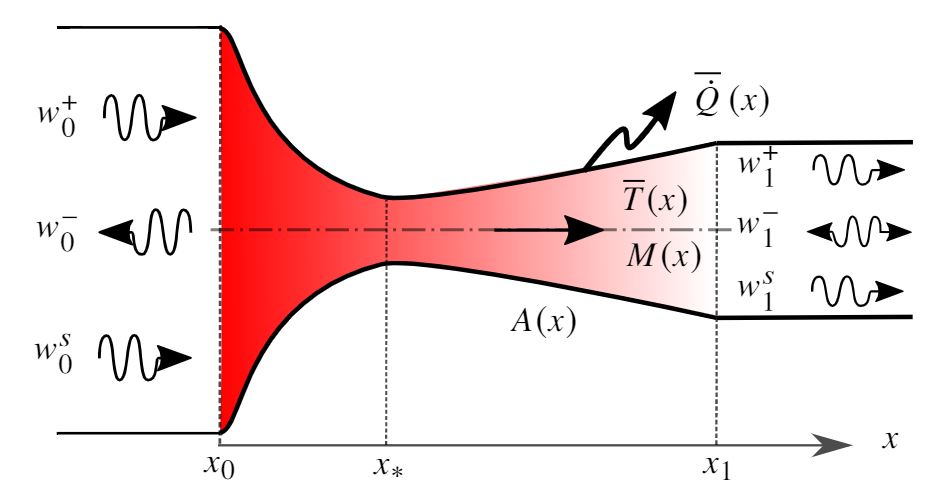}}
  \caption{Sketch of a nozzle of length $L$ exchanging heat with the surroundings. $A(x)$, $M\left(x\right)$, and $\overline{T}(x)$ represent the nozzle cross-sectional area, flow Mach number and mean-temperature at any $x$ varying from ${x}_{0}$ to ${x}_{1}$.}
\label{fig:Figure1}
\end{figure}
We consider a calorically perfect gas flowing through a nozzle as 
depicted in figure~\ref{fig:Figure1} with steady heat transfer which varies axially as $\overline{\dot{Q}}\left(x\right)$. The flow is taken to be inviscid and compressible, and a quasi-one-dimensional framework is adopted. Neglecting volumetric forces as well as thermal and mass diffusion, the conservation of mass, momentum, energy and equation of state can be written, respectively,
as,
\refstepcounter{equation}
$$
    A \frac{ \partial \rho  }{ \partial t } +\frac{\partial(\rho Au)}{\partial x} = 0, \quad     \frac{ \partial u }{ \partial t } +u\frac{\partial{u}}{\partial x} + \frac{1}{\rho}\frac{\partial p}{\partial x} = 0, \quad
    \frac{ \partial {s} }{ \partial t } +{u}\frac{\partial{s}}{\partial x}  = \frac{{R}_{g}}{p}{\overline{\dot{Q}}} \quad \text{and} \quad {p} = {\rho} {R}_{g} {T},
  \eqno{(\theequation{\mathit{a}~-~\mathit{d}})}\label{eq:ConservationEqs}
$$
 where ${\rho}$ is density, ${p}$ is pressure, ${T}$ is temperature, 
 ${u}$ is axial velocity, ${s}$ is entropy, ${R}_{g}$ is the gas constant, $A$ is the cross-sectional area of the nozzle and ${\overline{\dot{Q}}}$ denotes the steady volumetric heat source term.
 
We seek to retrieve the dynamics of unsteady, small-amplitude perturbations superimposed
on a steady background mean flow. The linearisation principle allows the thermodynamic and flow variables 
to be decomposed into the sum of mean time-averaged and fluctuating time-dependant components, denoted by $\overline{(\;\:)}$ and ${(\;)'}$, respectively, e.g  $\rho = \overline{\rho} + \rho'$. These time-dependant quantities are further normalised as follows,
 \refstepcounter{equation}
$$
\hat{\rho}=\dfrac{\rho'}{\overline{\rho}}, \quad \hat{p}=\dfrac{p'}{\gamma\overline{p}}, \quad \hat{u}=\dfrac{u'}{\overline{u}}, \quad  \hat{s}=\left(\gamma - 1\right)\dfrac{s'}{\gamma R_g}
  \eqno{(\theequation{\mathit{a}~-~\mathit{d}})}\label{eq:Normalisation}
$$
with $\gamma$ denoting the adiabatic index. It is assumed that there are no heat fluctuations inside the nozzle (${\dot{Q}}^{'} = 0$). The analysis could be extended to account for heat fluctuations, but would require an additional closure model linking these to flow fluctuations. For simplicity, this paper focuses on the primary effect of mean heat transfer.
  
By linearising the thermodynamic and flow variables in the conservation equations~\eqref{eq:ConservationEqs} and equating the mean quantities, we obtain:
 \refstepcounter{equation}
$$
    \dfrac{1}{\overline{\rho}}\dfrac{\textrm{d}\overline{\rho}}{{\textrm{d}x}}+ \dfrac{1}{\overline{u}}\dfrac{\textrm{d}\overline{u}}{\textrm{d}x} + \dfrac{1}{A}\dfrac{\textrm{d}A}{\textrm{d}x} = 0, \quad \overline{u}\dfrac{\textrm{d}\overline{u}}{{\textrm{d}x}} = -\dfrac{1}{\overline{\rho}}\dfrac{\textrm{d}\overline{p}}{\textrm{d}x},\quad
    \dfrac{\gamma R_g}{\gamma - 1}\dfrac{\textrm{d}\overline{T}}{\textrm{d}x} + \overline{u}\dfrac{\textrm{d}\overline{u}}{{\textrm{d}x}} = \dfrac{\overline{\dot{Q}}}{\overline{\rho} \:\overline{u}} \quad \textrm{and} \quad \overline{p} = \overline{\rho}R_g \overline{T}.
  \eqno{(\theequation{\mathit{a}~-~\mathit{d}})}\label{eq:MeanFlowEqs}
$$
Similarly, the linearised forms of the Euler equations are obtained in the time domain by equating the first order fluctuating quantities of the conservation equations~\eqref{eq:ConservationEqs} in their normalised form given by~\eqref{eq:Normalisation}. They read:
\begin{equation}\label{eq:NormalisedMassEnergy}
    \dfrac{ \partial \hat{p}  }{ \partial t } + {\overline{u}} \dfrac{\partial}{\partial x}\left(\hat{p} + \hat{u} \right) = - \overline{\dot{Q}} \dfrac{\gamma - 1}{\gamma \overline{p}}\left(\hat{u} + \gamma\hat{p}\right),
\end{equation}
\begin{equation}\label{eq:NormalisedXMomentum}
    \dfrac{ \partial \hat{u}}{ \partial t } + {\overline{u}} \dfrac{\partial \hat{u}}{\partial x} + \dfrac{\overline{c}^2}{\overline{u}} \dfrac{\partial \hat{p}}{\partial x} + \dfrac{\textrm{d}{\overline{u}}}{\textrm{d} x}\left[2\hat{u} - \left(\gamma-1\right)\hat{p} - \hat{s} \right] = 0,
\end{equation}
\begin{equation}\label{eq:NormalisedEntropy}
        \dfrac{ \partial \hat{s}}{ \partial t } + {\overline{u}} \dfrac{\partial \hat{s}}{\partial x} = - \overline{\dot{Q}} \dfrac{\gamma - 1}{\gamma \overline{p}}\left(\hat{u} + \gamma\hat{p}\right),
\end{equation}
with $\overline{c}$ denoting the adiabatic speed of sound. The above linearised Euler equations (LEEs)~\eqref{eq:NormalisedMassEnergy}~-~\eqref{eq:NormalisedEntropy} govern the perturbed flow in a nozzle sustaining a mean flow with steady heat transfer.

To employ the Magnus expansion, the linearised Euler equations need to be recast in terms of three variables that are invariants of the flow at zero frequency and with zero heat transfer. The normalised fluctuating mass flow rate ($I_A = \hat{m}$), stagnation temperature ($I_B = \hat{T}_t$) and entropy ($I_C = \hat{s}$) are chosen as the three invariants, where,
\refstepcounter{equation}
$$
\hat{m} \equiv \dfrac{m'}{\overline{m}}, \quad \hat{T}_t \equiv \dfrac{T'_t}{\overline{T}_t},
  \eqno{(\theequation{\mathit{a}~-~\mathit{c}})}\label{eq:FlowInvariants1}
$$
with $m = \rho A u$ and $T_t = T\left(1 + \dfrac{\gamma-1}{2}\left(\dfrac{u}{c}\right)^2\right)$. In terms of the primitive variables $\hat{p}, \hat{u}$ and $\hat{s}$, these invariants read:
 \refstepcounter{equation}\
$$
I_{A}=\hat{p}+\hat{u}-\hat{s}, \quad
I_{B}=(\gamma-1) \dfrac{M^{2} \hat{u}+\hat{p}+\dfrac{\hat{s}}{\gamma-1}}  {1+\dfrac{\gamma-1}{2} M^{2}}, \quad
I_{C}=\hat{s}, 
  \eqno{(\theequation{\mathit{a}~-~\mathit{c}})}\label{eq:FlowInvariants}
$$
where $M=\overline{u}/\overline{c}$ is the local Mach number of the flow. 
By defining a vector of invariants such that $\mathsfbi{I} = {[I_A \;I_B\; I_C]}^{\textrm{T}}$ and using $\zeta = {1+\dfrac{\gamma-1}{2} M^{2}},$ the linearised Euler equations can be recast in matrix form. The full derivation is outlined in \S \ref{sec:appendix} and the final equation reads,
\begin{equation}\label{eq:FinalPDE}
\dfrac{\partial}{\partial t} \mathsfbi{I} \:+\:\overline{u} \mathsfbi{E}_{x} \frac{\partial}{\partial x} \mathsfbi{I}\:+\: \dfrac{\left(\gamma -1\right) \overline{\dot{Q}}}{2\zeta \gamma \overline{p}}\mathsfbi{E}_{s} \mathsfbi{I} = 0, \quad \textrm{where} \quad \mathsfbi{E}_x=\left[\begin{array}{ccc}
1 & \dfrac{\zeta}{(\gamma-1) M^{2}} & -\dfrac{1}{(\gamma-1) M^{2}} \\
\dfrac{\gamma-1}{\zeta} & 1 & \dfrac{\gamma-1}{\zeta} \\
0 & 0 & 1
\end{array}\right],  
\end{equation}

\begin{equation}\label{eq:Matrix_finalPDE}
\mathsfbi{E}_s=\dfrac{1}{M^2 - 1}\left[\begin{array}{ccc}
{M^2\left(1-\gamma\right)-2}& \zeta \dfrac{\gamma+1}{\gamma-1} & -\gamma M^2 - \dfrac{2\gamma}{\gamma-1} \\
2\gamma\left(\gamma-1\right) + \dfrac{M^2\left(1-\gamma^2\right)}{\zeta} & -2\zeta\gamma + \dfrac{\gamma^2 M^2}{\gamma-1} & 2\gamma\left(\gamma-\dfrac{ M^2}{\zeta}\right) -\dfrac{\gamma\left(\gamma-1\right)}{\zeta}M^4\\
2\zeta\left(\gamma M^2-1\right) & -2{\zeta}^2 & 2\zeta\gamma M^2
\end{array}\right].
\end{equation}
As observed in~\eqref{eq:FinalPDE} and~\eqref{eq:Matrix_finalPDE}, the coefficients of the matrices are only functions of the axial mean flow and the mean rate of heat transfer per unit volume, $\overline{\dot{Q}}$. 

A harmonic time-dependence of the fluctuating quantities is now assumed, such that $\hat{y} = \breve{y}{\text{e}}^{\text{i}\omega t}$, with $\omega$ the angular frequency and $\textrm{i}^2={-1}$. Equation~\eqref{eq:FinalPDE} is recast in the frequency domain as, 
\begin{equation}\label{eq:Final1stOrderODE}
 \dfrac{\textrm{d}}{\textrm{d} x}  \mathsfbi{I}=\mathsfbi{A}{\left(x, \:\omega,\:\overline{\dot{Q}}\right)}\mathsfbi{I} \quad \textrm{with} \quad \mathsfbi{A}  = -{\left[\mathsfbi{E}_{x}\right]}^{-1}\left(\dfrac{\textrm{i}\omega}{\overline{u}}\mathcal{I} + \dfrac{\left(\gamma-1\right)\overline{\dot{Q}}}{2\zeta \gamma \overline{p}\: \overline{u}} \mathsfbi{E}_{s}\right),
\end{equation}
where $\mathcal{I}$ is an identity matrix of order 3. 
Equation~\eqref{eq:Final1stOrderODE} is similar to the equation obtained by~\cite{Duran_JFM_2013} but with an extra term, involving the matrix $\mathsfbi{E}_{s}$, which accounts for mean heat transfer effects.
The frequency and the heat source are specified in terms of the non-dimensional parameters $\Omega = \omega L/\overline{c}_{t0}$ (axial Helmholtz number) and $\tilde{Q} = \overline{\dot{Q}}L/\left(\overline{p}_{t0}\; \overline{c}_{t0}\right),$ respectively, where $L$ is the length of the nozzle, and $\overline{p}_{t0}$ and $\overline{c}_{t0}$ correspond to the stagnation pressure and stagnation speed of sound at the inlet, respectively.

Equation~\eqref{eq:Final1stOrderODE} is solved 
using a Magnus-expansion-based method, that assumes the ansatz,
\begin{equation}\label{eq:MagnusInvariantResult}
\mathsfbi{I}(\xi, \Omega, \tilde{Q})=[\exp [\mathsfbi{B}(\xi, \Omega, \tilde{Q})]] \mathsfbi{I}_{0} \quad \text { with } \quad \mathsfbi{B}(\xi, \Omega, \tilde{Q})=\sum_{k=0}^{\infty} \mathsfbi{B}^{(k)}(\xi, \Omega, \tilde{Q}),
\end{equation}
with $\xi$ = $x/L$ denoting the dimensionless axial coordinate and $\mathsfbi{I}_{0}$ the
vector of invariants at the inlet. $\mathsfbi{B}^{(k)}(\xi, \Omega, \tilde{Q})$ represents the terms of the Magnus expansion each of order $O(\Omega, \tilde{Q})^{k}.$ The terms in the expansion are obtained recursively as explained by \cite{Blanes_PR_2009}. When the frequency of the fluctuating components (represented by $\Omega$) and the heat exchange (represented by $\tilde{Q}$) are zero, the flow invariants defined in~\eqref{eq:FlowInvariants} are conserved and remain constant along the nozzle length. Any non-zero frequency and/or non-zero heat exchange results in a deviation which is captured by the higher order terms  ($k\neq0$) in the Magnus expansion. 
The main novelty of this study is that, 
for the first time, a term capturing flow non-isentropicity is included as the expansion parameter along with frequency. To ensure convergence, the series may require the separate computation of transfer matrices for axial segments of the nozzle, which are subsequently multiplied for the final result \citep{Blanes_PR_2009}. For example, for the isentropic sub-critical mean nozzle flow considered in \S\ref{sec:Results}, at low frequencies (${\Omega}/{2\pi} \sim 0.0001 - 0.01$) the series exhibits fast convergence when applied to the entire nozzle length and does not require any axial segmentation. However at higher frequencies, for example when ${{\Omega}}/2\pi = 1$, the nozzle needs to be divided into at-least 24 axial segments for fast convergence. This segmentation approach follows the fast convergence criterion given by,
\begin{equation}\label{eq:Convergence}
\int_{0}^{\xi_{F}}\|\boldsymbol{A}(\xi)\|_{2} \mathrm{~d} \xi<\pi,
\end{equation}
where $\xi_F$ determines the maximum length of the segments. The above equation~\eqref{eq:Convergence} ensures faster convergence when satisfied, but may diverge if the value of the integral is larger than $\pi$~(\citet{Blanes_PR_2009}).

Finally, the flow invariants at the boundaries are transformed into three propagating waves (see figure~\ref{fig:Figure1}): (i) a downstream-propagating acoustic wave, $w^+$, (ii) an upstream-propagating acoustic wave, $w^-,$ and (iii) an entropy wave, $w^s.$ These wave components at a particular location are represented using the wave vector $\mathsfbi{W}=\left[w^+ \;w^-\; w^s\right]^{\rm T}$. In terms of the primitive variables, they read $w^+ = \breve{p} + \breve{u}M$, $w^- = \breve{p} - \breve{u}M,$ and $w^s = \breve{s}.$

The relation between the flow invariants and the wave components can then be obtained using \eqref{eq:FlowInvariants} and is represented as $\mathsfbi{I} = \mathsfbi{D}\mathsfbi{W}$. A transfer matrix $\mathsfbi{T}$ is therefore defined to relate the wave vector at any location, $\xi$, in terms of the wave vector at the inlet ($\mathsfbi{W}_{0}$) as, 
\begin{equation}\label{eq:TransferMatrix}
\mathsfbi{W}_{\xi} =  \mathsfbi{T} \mathsfbi{W}_{0} \quad \textrm{where} \quad \mathsfbi{T} =  \left[\mathsfbi{D}_{\xi}\right]^{-1} \mathsfbi{C} \; \mathsfbi{D}_{0},
\end{equation}
and $\mathsfbi{C} = [\exp [\mathsfbi{B}(\xi, \Omega, \tilde{Q})]]$.
\subsection{Sub-critical nozzle flow configuration}
For the case where the flow remains subsonic inside the nozzle,
the acoustic system is determined by three external  inputs, namely a downstream-propagating acoustic wave at the inlet, $w^+_{0,\:f}$, an upstream-propagating acoustic wave at the outlet, $w^-_{1,\:f},$ and an 
entropy wave at the inlet, $w^s_{0,\:f}$. Note that the subscripts `0' and `1' refer to the inlet and outlet of the nozzle, respectively, while `$f$' denotes that the wave is externally forced. We also have three waves as the outputs of the system, i.e. an upstream-propagating acoustic wave at the inlet, $w^-_0$, a downstream-propagating acoustic wave at the outlet, $w^+_{1}$ and an entropy wave at the outlet, $w^s_{1}$.  An extended scattering matrix $\mathsfbi{S}$ can then be defined to relate the incoming and outgoing wave vectors as,
\begin{equation}\label{eq:ScatteringMatrixSubsonic}
    {\left[\begin{array}{c}
w^+_1 \\w^-_0 \\ w^s_1 \end{array}\right]}\; = \mathsfbi{S} {\left[\begin{array}{c}
w^+_{0,\:f} \\w^-_{1,\:f} \\ w^s_{0,\:f} \end{array}\right]}.
\end{equation}
This matrix is obtained by algebraically rearranging the terms of the matrix $\mathsfbi{T}$ in \eqref{eq:TransferMatrix}.
\subsection{Super-critical nozzle flow without shocks}
For a super-critical flow without any shock-waves inside the nozzle, only two inputs, namely $w^+_{0,\:f}$ and $w^s_{0,\:f},$ can be applied as $w^-_{1}$ now corresponds to the slow downstream propagating wave at the nozzle outlet. The flow domain is then divided into subsonic (from the inlet $x_0$ to a location infinitesimally upstream of the throat $x_u$) and supersonic (from a location infinitesimally 
downstream of the throat $x_d$ to the outlet $x_1$) portions. Mass fluctuations cannot occur at the choked throat and this requirement is specified as a boundary condition~\citep{Marble_JSV_1977} which takes the form ${M}^{'}/M = 0$. At the location infinitesimally upstream of the throat, this gives
\begin{equation}\label{eq:ThroatCondition}
   w^-_u = \cancelto{0}{w_{u,\:f}^-} \; + R_u w^+_u + R_s w^s_u, \quad \textrm{with} \quad R_u = \dfrac{3-\gamma}{1+\gamma}, R_s = \dfrac{-2}{1+\gamma}
\end{equation}
where $R_u$ and $R_s$ correspond to the acoustic and entropy reflection coefficients, respectively. The inlet acoustic and entropy forcing inputs, $w^+_{0,\:f}$ and $w^s_{0,\:f},$ respectively, along with \eqref{eq:ThroatCondition} give the three input conditions. Equation \eqref{eq:TransferMatrix} then takes the form,
\begin{equation}\label{eq:TransferMatrixSupercriticalSub}
    {\left[\begin{array}{c}
w^+_u \\ {w^-_{u,\:f}} \\ w^s_u \end{array}\right]}\;\equiv\;{\left[\begin{array}{c}
w^+_u \\ 0 \\ w^s_u \end{array}\right]}\;=\underbrace{{\left[\mathsfbi{D}_{1}^{'}\right]}^{-1} \mathsfbi{C}_{\textrm{sub}} \; \mathsfbi{D}_{0}}_{{\mathsfbi{T}}^{'}} {\left[\begin{array}{c}
w^+_{0,\:f}\\ w^-_{0} \\ w^s_{0,\:f}\end{array}\right]}.
\end{equation}
Again rearranging the matrices with forcing inputs on one side and unknowns on the other gives,
\begin{equation}\label{eq:TransferMatrixSupercriticalSub_New}
    {\left[\begin{array}{c}
w^+_u \\ w^-_0 \\ w^s_u \end{array}\right]}\; = \mathsfbi{S}_{s} {\left[\begin{array}{c}
w^+_{0,\:f}\\ 0 \\ w^s_{0,\:f}\end{array}\right]} \Rightarrow w^-_0 = S_s(2, 1)\;w^+_{0,\:f} + S_s(2, 3)\;w^s_{0,\:f}.
\end{equation}
where $\mathsfbi{S}_{s}$ corresponds to the scattering matrix of subsonic portion of the super-critical nozzle, given in terms of the elements of $\mathsfbi{T}^{'}$.
Separating the upstream propagating wave at the inlet using \eqref{eq:TransferMatrixSupercriticalSub_New} gives,
\begin{equation}\label{eq:TransferMatrixSupercriticalSub_New2}
    {\left[\begin{array}{c}
w^+_u \\ 0 \\ w^s_u \end{array}\right]}\; = {\mathsfbi{S}'_{s}} {\left[\begin{array}{c}
w^+_{0,\:f}\\ 0 \\ w^s_{0,\:f}\end{array}\right]} \quad \text{where} \quad
{\mathsfbi{S}'_{s}} = \left[\begin{array}{c c c}
 S_s(1, 1)\quad 0 \quad S_s(1, 3)\\ 0 \quad 0 \quad 0 \\S_s(3, 1) \quad 0 \quad S_s(3, 3) \end{array}\right].  
\end{equation}
The wave vector is unaltered at the throat and hence $\mathsfbi{W}_{u} = \mathsfbi{W}_d$. Note that $\mathsfbi{W}_{u} = \left[w^+_{u} \; 0 \; w^s_{u}\right]^{\rm T}$ while $\mathsfbi{W}_{d} = \left[w^+_{d} \; {w^-_{d}} \; w^s_{d}\right]^{\rm T}$ since $w^-_{u,\:f} = 0$. Denoting the transfer matrix in the supersonic portion of the nozzle by $\mathsfbi{T}_{s}$, the wave vector relation $\mathsfbi{W}_{1}= \mathsfbi{T}_{s} \mathsfbi{W}_{d}$ is obtained similar to ~\eqref{eq:TransferMatrix}, where $\mathsfbi{T}_{s} = {\left[\mathsfbi{D}_{1}\right]}^{-1} \mathsfbi{C}_{\textrm{super}} \; \mathsfbi{D}_{d}$. This gives:
\begin{equation}\label{eq:TransferMatrixSupercritical}
{\left[\begin{array}{c}
w^+_1 \\ w^-_1 \\ w^s_1 \end{array}\right]}\; = \mathsfbi{T}_{s}
{\mathsfbi{S}'_{s}} {\left[\begin{array}{c}
w^+_{0,\:f}\\ 0 \\ w^s_{0,\:f}\end{array}\right]}.
\end{equation}            
Equation \eqref{eq:TransferMatrixSupercritical} gives the fast ($w^+_{1}$) and slow ($w^-_{1}$) propagating acoustic waves and the entropy wave ($w^s_{1}$) at the outlet as a function of acoustic ($w^+_{0,\:f}$) and entropy ($w^s_{0,\:f}$) forcing at the nozzle inlet. The upstream propagating wave at the inlet ($w^-_{0}$) in the subsonic portion is given by \eqref{eq:TransferMatrixSupercriticalSub_New}.

\section{Results}\label{sec:Results}
In this section, we apply the presented theory to a converging-diverging 
nozzle (see figure~\ref{fig:Figure1}) defined by the following area profile  
\begin{equation}
    \frac{A(x)}{A_{*}}=\left\{\begin{array}{ll}
\dfrac{1}{2}\left(\dfrac{A_{\text{0}}}{A_{*}}-1\right)\left[\cos \left(\upi \dfrac{x}{x_{*}}\right)+1\right]+1 & \text { if } x \in\left[0, x_{*}\right] \\
1+\left(\dfrac{A_{\text{1}}}{A_{*}}-1\right) \dfrac{x-x_{*}}{L-x_{*}} & \text { if } x \in\left[x_{*}, L\right]
\end{array}\right.
\end{equation}
where $A_* = 0.002 \textrm{m}^2$ corresponds to the 
throat area (at the throat location $x_* = 0.15$m), $A_{0}/{A_*} =  2.1$
and $A_{1}/{A_*} = 1.18$. The gas constant and adiabatic index are taken to be ${R}_{g}=287 \text{J} \text{kg}^{-1} \text{K}^{-1}$ and $\gamma=1.4,$ respectively.

\begin{figure}
\begin{subfigure}{.5\textwidth}
  \centering
  \includegraphics[width=\linewidth]{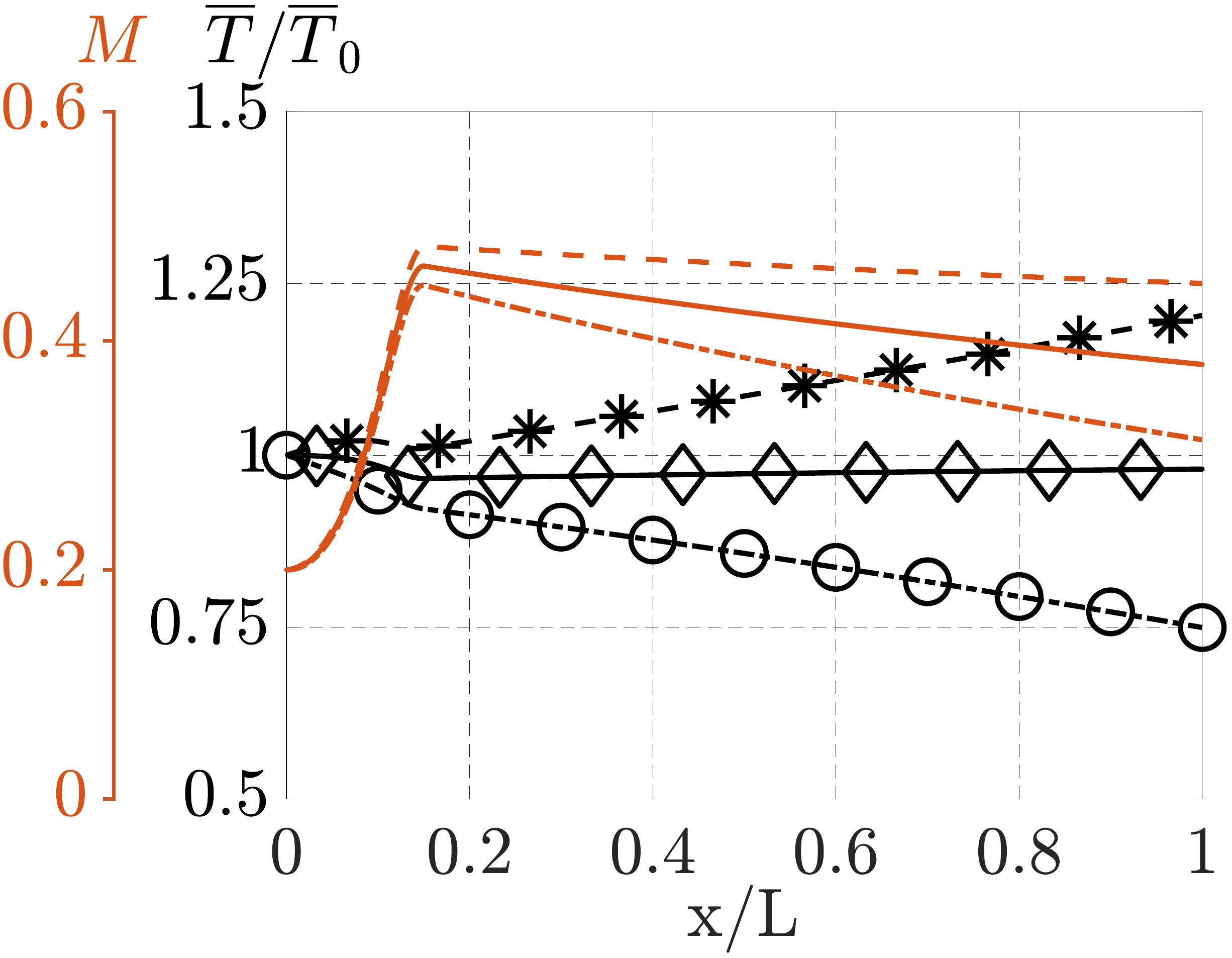}
  \put (-195,150) {\normalsize$\displaystyle(a)$}
    \vspace*{-0pt}
  \label{fig:sFigure2a}
\end{subfigure}%
\hspace{4pt}
\begin{subfigure}{.5\textwidth}
  \centering
  \includegraphics[width=\linewidth]{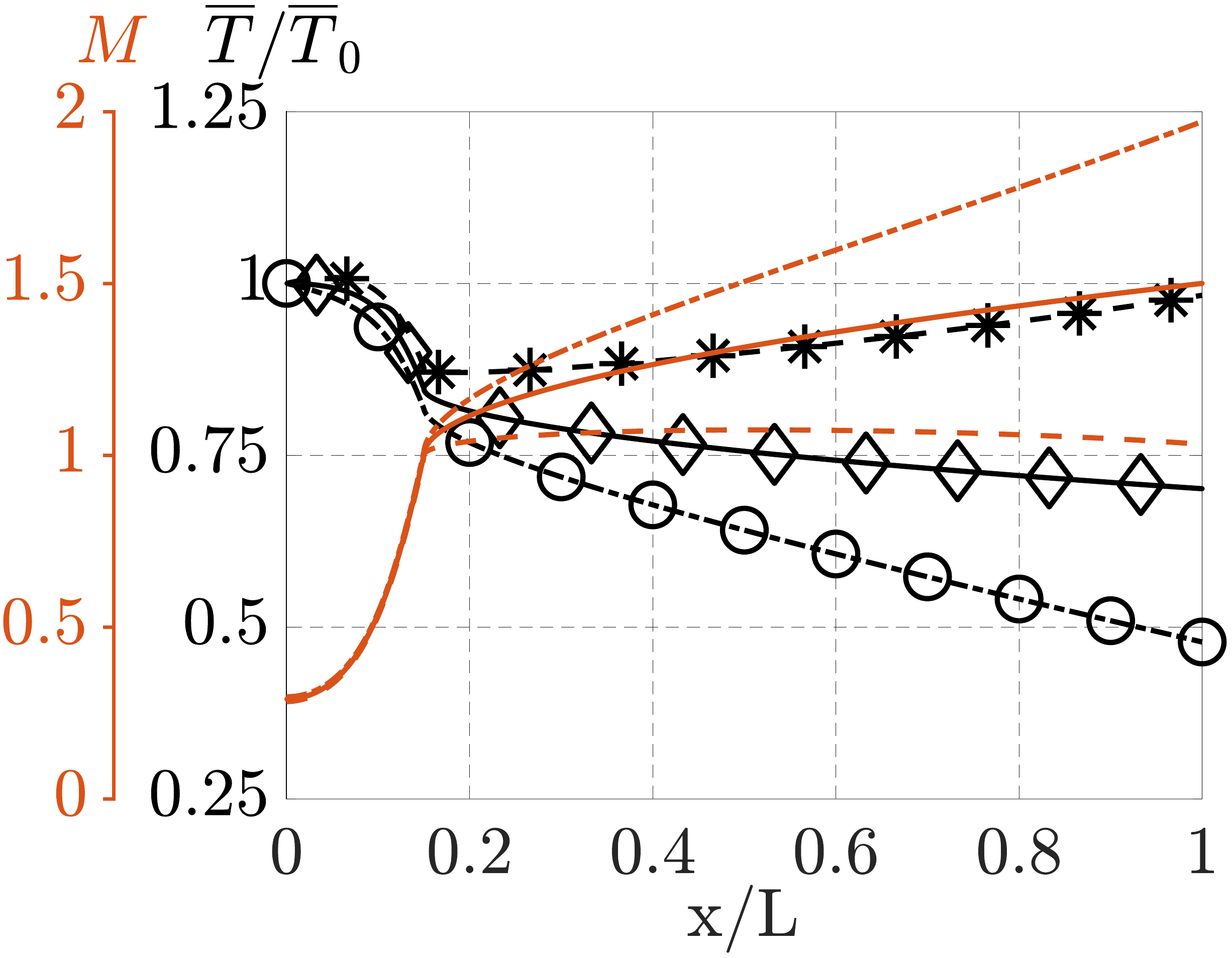}
  \put (-195,150) {\normalsize$\displaystyle(b)$}
   \vspace*{-0pt}
  \label{fig:sFigure2b}
\end{subfigure}
\vspace{-0.6cm}
\caption{Mach number, M, (lines) and mean-temperature, $\overline{T}$, (lines with markers) for: ($a$) sub-critical case with $\tilde{Q}$ values of -0.5~(\chain), 0~(-----), 0.5~(\dashed); ($b$) super-critical case for $\tilde{Q}$ values of -0.5~(\chain), 0~(-----), 0.3~(\dashed).}
\label{fig:Figure2}
\end{figure}

Two different flow cases are considered: (i) sub-critical and 
(ii) super-critical (without shocks).  For both flow cases, numerical solutions of the mean flow are obtained for different levels of heat transfer, set by $\tilde{Q}.$ In the following, we assume for simplicity $\tilde{Q}$ to be constant axially, but the approach can assume any spatial distribution. The sub-critical case is defined by an inlet Mach number of $M_{0} = 0.2.$ The super-critical one assumes $M_{0} = 0.29$ for the isentropic case. When heat is added the inlet Mach number ranges from $M_0=0.28$ to $M_0=0.30$ for $\tilde{Q}=0.3$ and $\tilde{Q}=-0.5,$ respectively. The mean flow for the sub-critical case is obtained solving the first order system of differential equations of \eqref{eq:MeanFlowEqs} using higher order implicit schemes: for example, a fourth order Runge-Kutta was used here. For the super-critical case, a modified finite difference based MacCormack scheme is used to compute the mean flow. The evolution of the Mach number and mean temperature inside the nozzle are presented in figure \ref{fig:Figure2}. As expected, the temperature at the outlet is higher when heat is added than for the isentropic flow. For the Mach number distribution, we observe different trends for the sub-critical and super-critical cases. For the sub-critical nozzle flow, the Mach number at the outlet increases when heat is added owing to the decrease of density and subsequent increase of the flow velocity to satisfy continuity. For the super-critical nozzle flow, adding heat reduces the Mach number at the outlet, explained by the increase of the speed of sound due to the temperature rise.
\subsection{Effect of heat transfer at zero frequency}
\begin{figure}
	\centering
	\includegraphics[width=\textwidth]{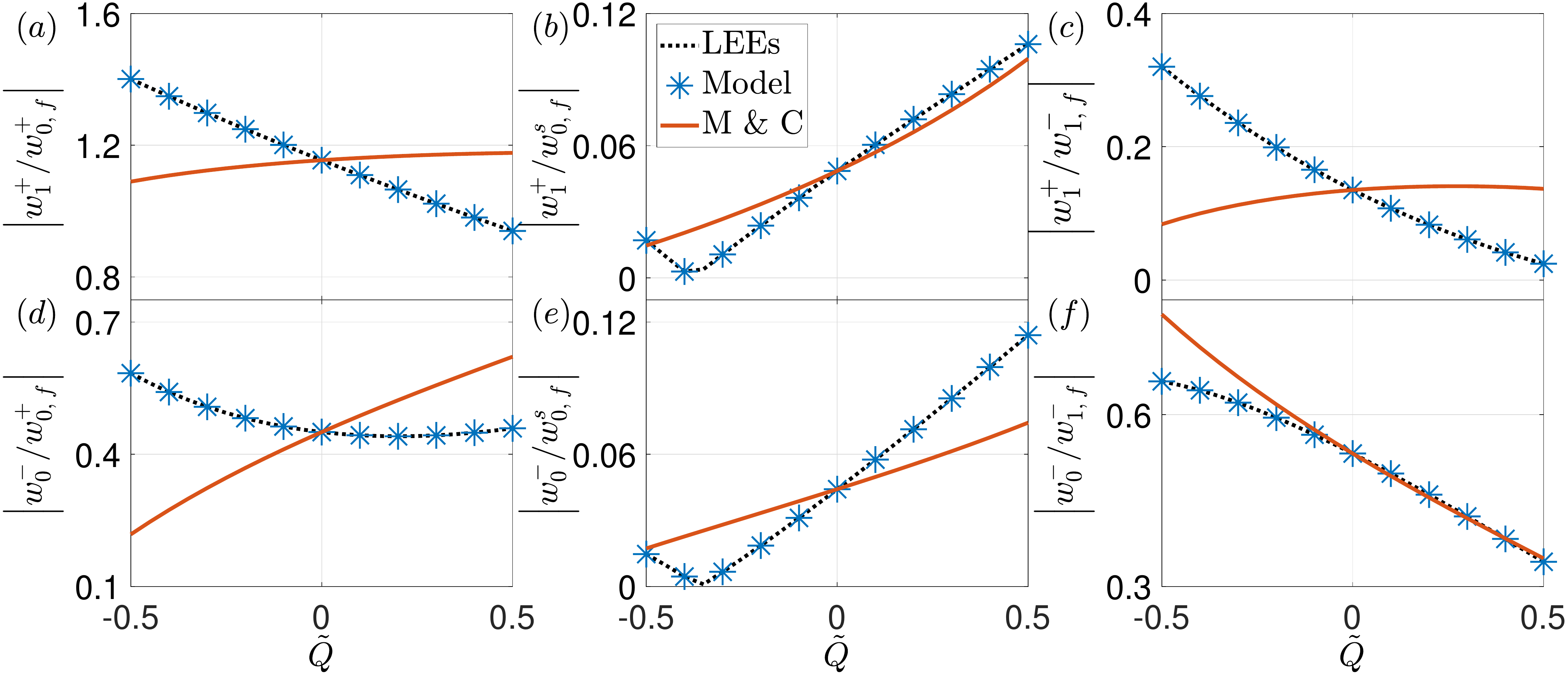}
  \caption{Transfer functions for the sub-critical nozzle flow at zero frequency predicted by the present model~(\textcolor{MATblue}{\textasteriskcentered}), the compact isentropic model of \cite{Marble_JSV_1977}~(\textcolor{MATred}{----}), and numerical solutions (\dotted).}
	 \label{fig:Figure3}
	 \vspace*{00pt}
\end{figure}
\begin{figure}
	\centering
	\includegraphics[width=\textwidth]{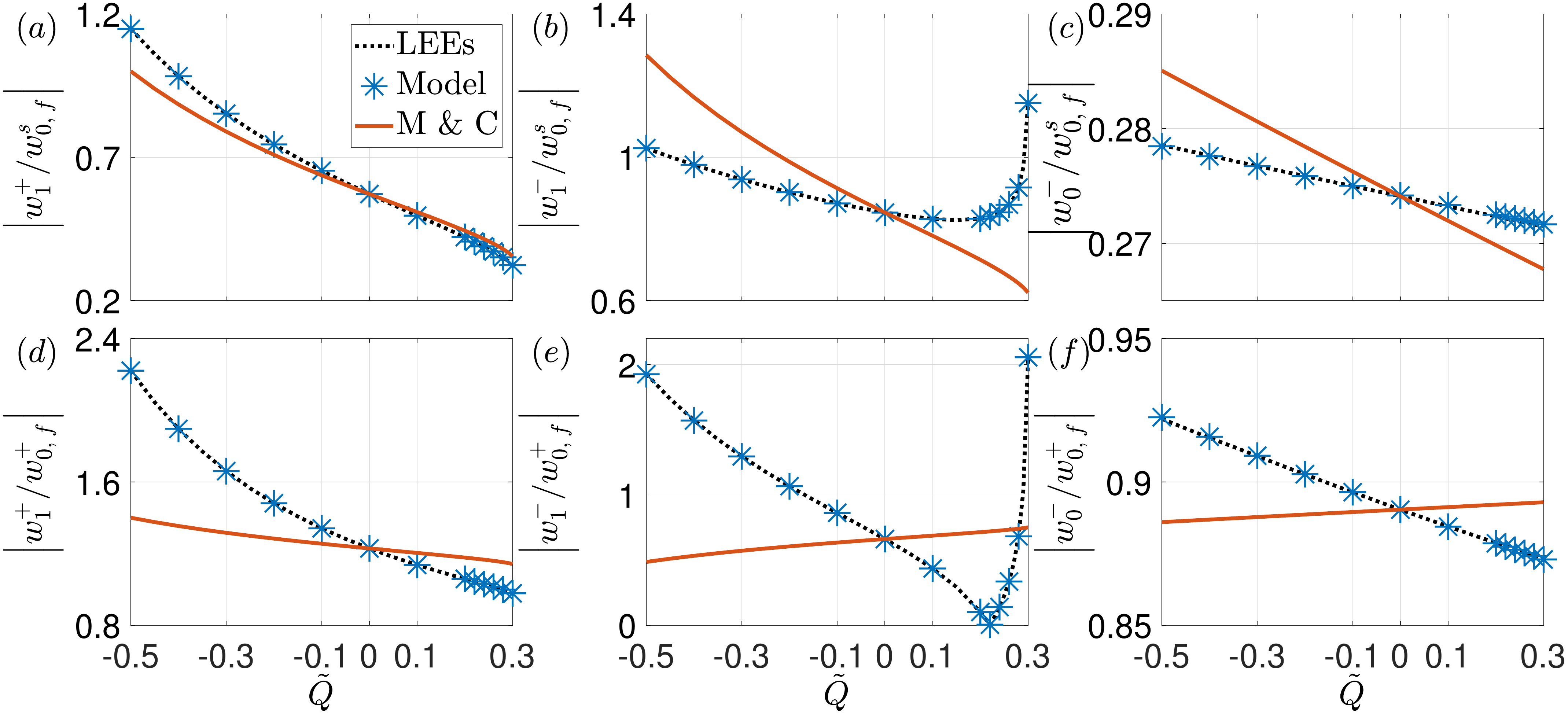}
   \caption{Transfer functions for the super-critical nozzle flow at zero frequency 
   predicted by the present model~(\textcolor{MATblue}{\textasteriskcentered}), 
   the compact isentropic model of~\cite{Marble_JSV_1977}~(\textcolor{MATred}{----}), and numerical solutions~(\dotted).}
	 \label{fig:Figure4}
\end{figure}
In this section, we explore the influence of heat transfer on the 
unsteady response of the nozzle at zero frequency ($\Omega = 0)$. The heat source/sink is varied in the range $\tilde{Q}\in \left[-0.5, 0.5\right]$ for the sub-critical flow and $\tilde{Q}\in \left[-0.5, 0.3\right]$ for the super-critical one.
For $\tilde{Q}\gtrsim 0.3$ in the super-critical regime, a normal shock appears. While the model can be extended to nozzles sustaining normal shocks~(\cite{Duran_JFM_2013}), the current validation is limited to shock-free nozzles only.
Figures~\ref{fig:Figure3} and~\ref{fig:Figure4} show the coefficients of the acoustic transfer functions for the sub-critical and super-critical nozzle flow cases, respectively.
The order of the Magnus expansion considered is $k=5$. 
The model predictions are compared with numerical solutions and
an excellent agreement is observed. 

To assess the importance of the heat transfer in the acoustic solution, the isentropic and compact theory of~\cite{Marble_JSV_1977} is also plotted. This model requires only two parameters, namely the Mach number at the inlet and at the outlet. Here, we feed those parameters to the Marble and Candel model based on the mean flow numerical solution with heat transfer. This results in predictions where the perturbations are 
isentropic, but the mean flow is corrected to account for non-isentropicity. As observed, the effect of heat transfer on the results is strong: the trends for some of the coefficients are reversed when compared with isentropic theory and the relative errors are as high as several hundred percent. The differences can be explained based on the governing equations of the fluctuating variables. The isentropic and non-isentropic equations only differ in the source term owing to heat transfer on the right-hand side of~\eqref{eq:NormalisedMassEnergy} and~\eqref{eq:NormalisedEntropy}. These terms represent sources of 
fluctuating mass flow rate, stagnation temperature and entropy that are produced by the interaction of acoustic waves with the steady heat transfer. The variations of fluctuating mass flow rate and stagnation temperature act as an additional source of sound. The variations of entropy, on the other hand, act as a dipole source of sound through the momentum equation~\eqref{eq:NormalisedXMomentum}. For the sub-critical case, the dominant effect is the former. This can be argued based on the the relatively low values of the acoustic response to entropy fluctuations (figures~\ref{fig:Figure3}(b) and (e)).
For values of around $\tilde{Q}\approx-0.4,$ this is especially notable, as the sound produced by entropy tends to zero, but the acoustic coefficients substantially differ from the predictions of isentropic theory. For the super-critical configuration, on the other hand, both effects are similarly important, as can be deduced from the high values of entropy-sound generation (figures~\ref{fig:Figure4}(a) and (b)). An interesting trend is 
observed for the slow-propagating downstream wave, $w^-_1,$ for the super-critical case: for values $0.1\lesssim \tilde{Q} \lesssim0.3,$ the coefficient exhibits a strong sensitivity when forced by either upstream acoustic waves or entropy waves. This effect, not observed for the fast-propagating downstream wave, may be linked to the fact that the Mach number is close to unity along the whole diverging portion of the nozzle in this range and, therefore, the propagation speed of that wave close to zero.

\subsection{Effect of heat transfer for non-zero frequencies}
\begin{figure}
	\centering
	\includegraphics[width=0.97\textwidth]{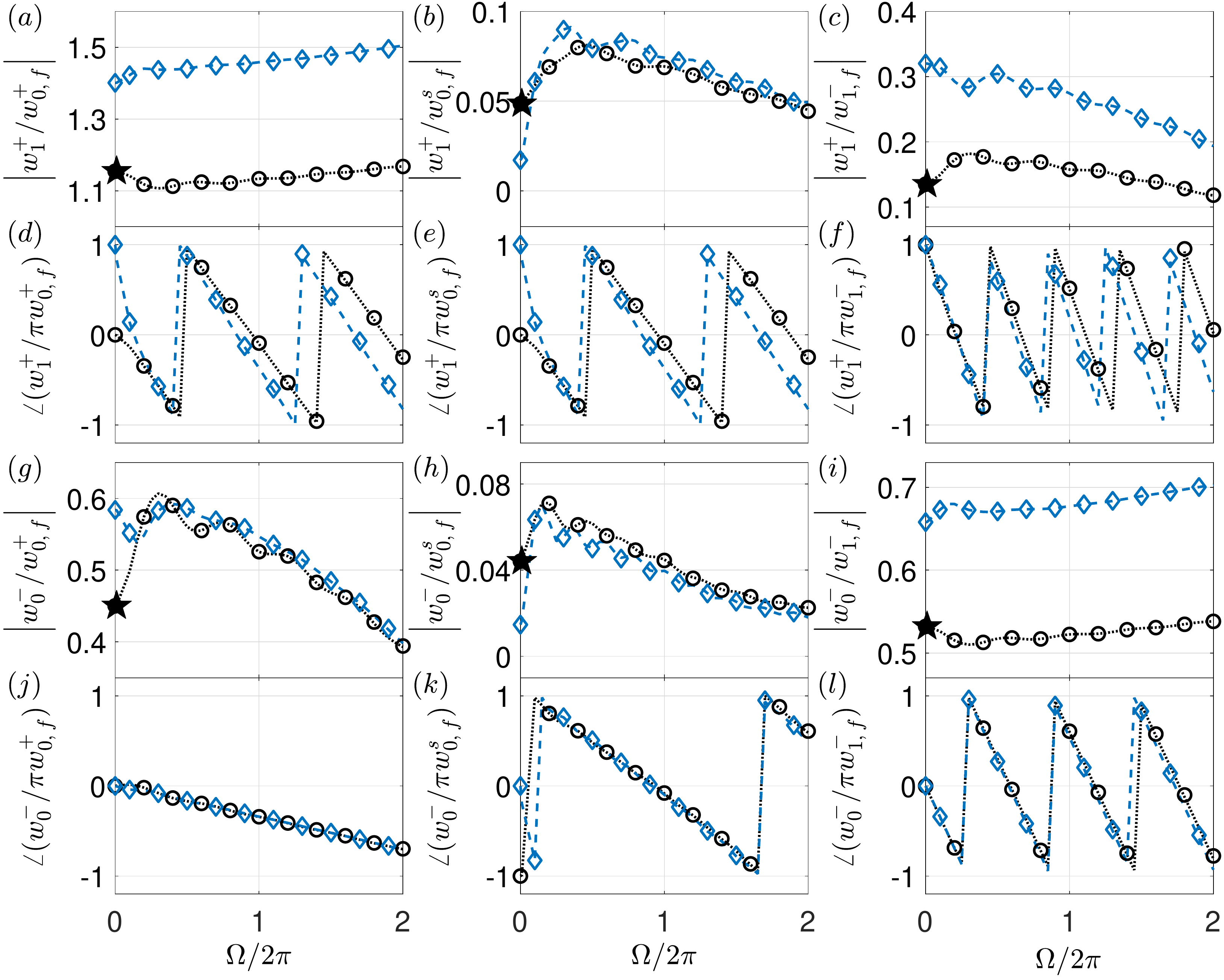}
 	\caption{Transfer functions for the sub-critical nozzle flow as a function of the frequency $\Omega$. Numerical (\dotted) and model solutions (\textbigcircle) for $\tilde{Q} = 0$. Numerical (\textcolor{MATblue}{\dashed}) and model solutions $\left(\textcolor{MATblue}{\lozenge}\right)$ for $\tilde{Q} = -0.5$. \cite{Marble_JSV_1977} compact solution $\left(\bigstar\right)$.}
	 \label{fig:Figure5}
	 \vspace*{00pt}
\end{figure}

We now turn our attention to the effect of heat transfer at non-zero frequencies. Figure~\ref{fig:Figure5} shows the variation of magnitude and phase of the acoustic transfer functions for the sub-critical nozzle flow, for both isentropic ($\tilde{Q}=0$) and non-isentropic ($\tilde{Q}=-0.5$) cases. It can again be observed that the model estimates closely match the numerical predictions. Similar trends are observed for all of the coefficients in the frequency domain. On comparing the isentropic and non-isentropic results for the magnitude of some of the coefficients, i.e. $w^+_1/w^+_{0,f},$ $w^+_1/w^-_{1,f},$ and $w^-_0/w^-_{1,f}$, an offset was observed at zero frequency which is carried to higher frequencies. For the rest of the coefficients, large differences exist at low frequencies, but quickly disappear with increasing frequencies. The phases for all of the coefficients are remarkably similar, with some differences observed for the coefficients defined in the downstream duct. This difference is expected due to the large difference in the flow Mach numbers and speed of sound in the divergent portion of the nozzle for nozzles with different levels of $\tilde{Q}$. This leads to different propagation velocities of the acoustic waves in the diverging portion of the nozzle and therefore to the differences in phase.
\begin{figure}
	\centering
	\includegraphics[width=0.7\textwidth]{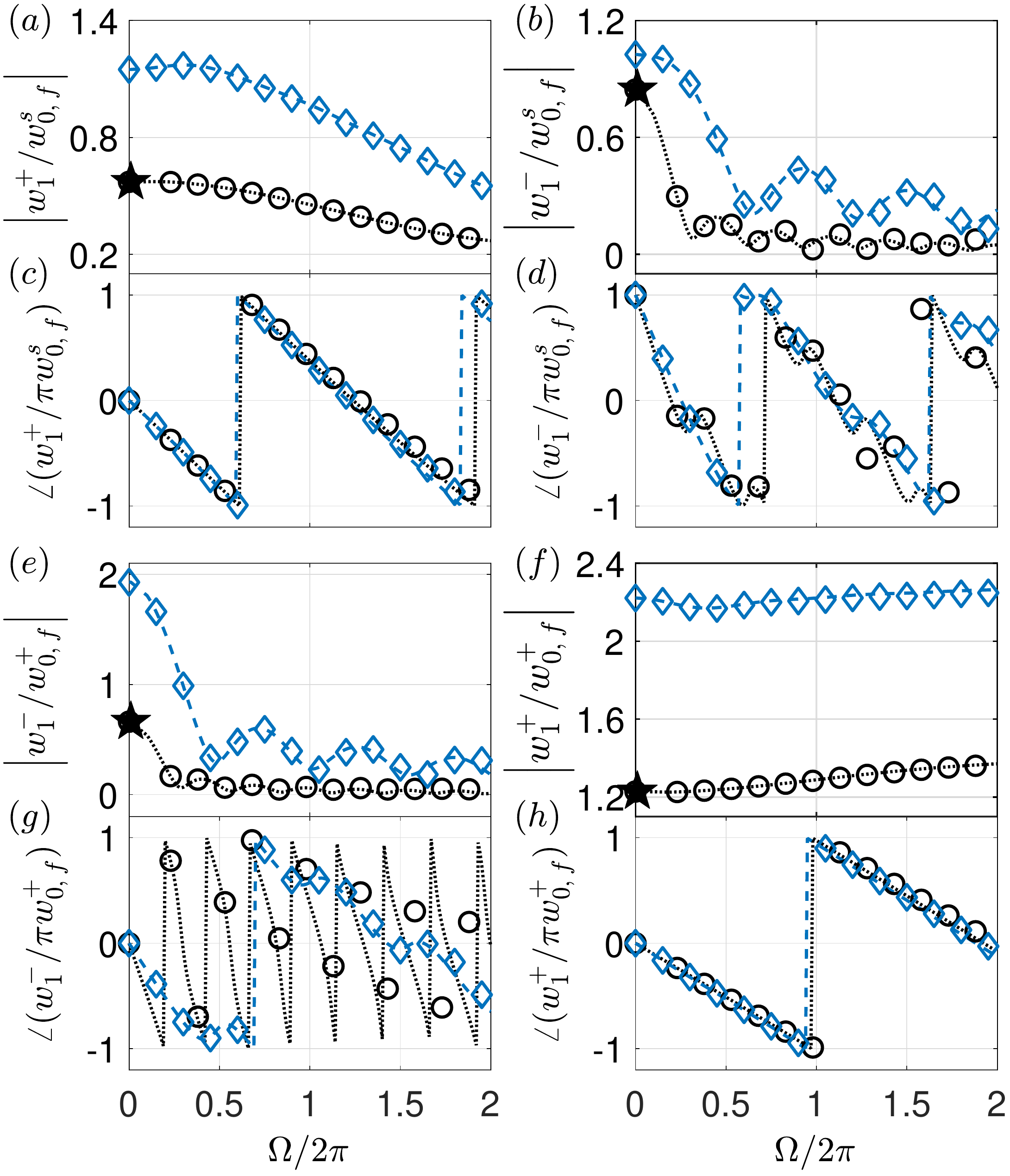}
 	 \caption{Transfer functions for the super-critical nozzle flow as a function of the frequency $\Omega$. Numerical (\dotted) and model solutions (\textbigcircle) for $\tilde{Q} = 0$. Numerical (\textcolor{MATblue}{\dashed}) and model solutions $\left(\textcolor{MATblue}{\lozenge}\right)$ for $\tilde{Q} = -0.5$. \cite{Marble_JSV_1977} compact solution $\left(\bigstar\right)$.}
	 \label{fig:Figure6}
	 \vspace*{00pt}
\end{figure}

Figure \ref{fig:Figure6} shows the variation of the acoustic transfer functions as a function of the frequency for the super-critical nozzle flow with $\tilde{Q} = -0.5$ and $0$. 
The model estimates are accurate for both cases.
The trends obtained for the reflection coefficient in the upstream duct $w^-_0$ are very similar to those obtained for the sub-critical case and have been omitted for the sake of brevity. This is consistent with $w^-_0$ not depending on the supersonic portion of the nozzle such that it can be obtained using \eqref{eq:ScatteringMatrixSubsonic} alone. The acoustic reflection at the outlet ($w^-_{1}$) has a low pass like behavior for both entropy ($w^s_{0,\:f}$) and acoustic ($w^+_{0,\:f}$) forcing at the inlet.

\section{Conclusions}\label{sec:Conclusions}
This work proposed a model for the acoustic and entropic transfer functions of non-isentropic nozzle flows where the non-isentropicity arises due to a steady source/sink of heat. Physically, this source term can be produced by heat exchange with surroundings or by chemical reactions. The flow was modelled as a quasi-one-dimensional and inviscid flow. A solution method was proposed based on the Magnus expansion, which does not impose any limitations on the frequency or Mach number that can be considered. In this work, a measure of the flow non-isentropicity has been used as an expansion parameter for the first time. The solution was successfully validated against numerical simulations of the quasi-one-dimensional linearised Euler equations in a converging-diverging nozzle for different heat transfer and frequency levels. Two types of nozzle flow were considered: sub-critical and super-critical (without shocks). For the acoustically compact case, it was shown that applying~\cite{Marble_JSV_1977} solution in the presence of heat transfer can lead to large prediction errors, even when the corrected mean flow Mach numbers are provided. The heat transfer was found to strongly affect the transfer functions of the nozzle, with some acoustic transfer functions being amplified and others attenuated, for increasing heat transfer.

\begin{acknowledgments}
\noindent
\textbf{Acknowledgments}: This work was supported by the European Research Council (ERC) Consolidator Grant AFIRMATIVE (2018-2023) and the Inlaks Shivdasani foundation.
\end{acknowledgments}

\noindent
\textbf{Declaration of interests:} The authors report no conflict of interest.

\noindent
\textbf{Author ORCIDs:}
\newline
\noindent
Saikumar Reddy Yeddula 
\url{https://orcid.org/0000-0002-1521-8948}

\noindent
Juan Guzmán-Iñigo
\url{https://orcid.org/0000-0002-1833-6034}

\noindent
Aimee S. Morgans \url{https://orcid.org/0000-0002-0482-9305} 

\appendix
\section{Linearised Euler equations in terms of flow Invariants}\label{sec:appendix}
Using the definition of flow invariants from~\eqref{eq:FlowInvariants} we can write, 
\begin{equation}\label{eq:DIPMatrix}
\overbrace{\left[\begin{array}{c}
I_A \\ I_B \\ I_C \end{array}\right]}^{\mathsfbi{I}}
 =\overbrace{\left[\begin{array}{cccc}
 1 & 1 & -1 \\
\dfrac{\gamma-1}{\zeta} & \dfrac{(\gamma-1)M^2}{\zeta}  & \dfrac{1}{\zeta} \\
0 & 0 & 1 \\
\end{array}\right]}^{\mathsfbi{D}_{I}^{P}}
\overbrace{\left[\begin{array}{c}
 \hat{p} \\ \hat{u} \\ \hat{s} \end{array}\right]}^{\mathsfbi{P}}, \quad \Rightarrow   \mathsfbi{P}=\left[\mathsfbi{D}_{I}^{P}\right]^{-1} I.
\end{equation}
By using the time averaged mean flow conservation equations~\eqref{eq:MeanFlowEqs}, we can write,
\begin{equation}\label{eq:SimpleMach}
    \dfrac{\textrm{d}\overline{u}}{\textrm{d}x} = \dfrac{\overline{u}}{\zeta}\left(\dfrac{1}{M}\dfrac{\textrm{d}M}{\textrm{d}x} + \dfrac{\left(\gamma-1\right)\overline{\dot{Q}}}{2 \gamma \overline{p}\:\overline{u}}\right).
\end{equation}
\begin{itemize}
\item \textbf{Invariant formulation for normalised fluctuating mass, $I_A$:}
    
On adding and subtracting~\eqref{eq:NormalisedXMomentum} and~\eqref{eq:NormalisedEntropy}, respectively, from ~\eqref{eq:NormalisedMassEnergy} and using~\eqref{eq:SimpleMach} gives,
\begin{equation}
    \begin{aligned}
        \dfrac{ \partial}{ \partial t }\left(\hat{p} +\hat{u}-\hat{s}\right) + \overline{u}\dfrac{ \partial}{ \partial x }\left(\hat{p} +\hat{u}-\hat{s}\right) - \dfrac{ \partial\hat{u}}{ \partial t } + \cancel{\left(\dfrac{ \partial\hat{s}}{ \partial t } + \overline{u}\dfrac{ \partial\hat{s}}{ \partial x }\right)} = \cancel{ \dfrac{-\left(\gamma-1\right)}{\gamma \overline{p}} \overline{\dot{Q}}\left(\hat{u} + \gamma \hat{p}\right)}.
    \end{aligned}
\end{equation}
Using~\eqref{eq:NormalisedXMomentum} we substitute for $\dfrac{ \partial\hat{u}}{ \partial t }$ and~\eqref{eq:SimpleMach} is used to replace $\dfrac{\text{d}\overline{u}}{\text{d}{x}}$. This gives,
\begin{equation}
    \begin{aligned}
        \dfrac{ \textrm{D} I_A}{ \textrm{D} t } + \dfrac{\overline{u}}{\zeta} \left(\dfrac{1}{M}\dfrac{\textrm{d}M}{\textrm{d}x} + \dfrac{\left(\gamma-1\right)\overline{\dot{Q}}}{2 \gamma \overline{p}\:\overline{u}}\right)\left(2\hat{u}-\left(\gamma-1\right) \hat{p}-\hat{s}\right) + \overline{u}\dfrac{\partial \hat{u}}{\partial x} + \dfrac{\overline{c}^2}{\overline{u}}\dfrac{\partial \hat{p}}{\partial x} = 0,
    \end{aligned}
\end{equation}
where $\dfrac{{\textrm{D}} {\left(\;\right)}}{{\textrm{D}} t} = \dfrac{\partial\left(\;\right)}{\partial t} + \overline{u}\dfrac{\partial\left(\;\right)}{\partial x}$. 
Taking the factor $1/M^2$ and writing out all the terms explicitly results in,
\begin{equation}
    \begin{aligned}
        &\dfrac{ \textrm{D} I_A}{ \textrm{D} t } + \dfrac{1}{M^2}\bigg(\bigg.2M\dfrac{\text{d}M}{\text{d}x}\dfrac{\overline{u}}{\zeta} \hat{u} - \left(\gamma - 1\right)M\dfrac{\text{d}M}{\text{d}x}\dfrac{\overline{u}}{\zeta}\hat{p} - M\dfrac{\text{d}M}{\text{d}x} \dfrac{\overline{u}}{\zeta}\hat{s}  + \\& \dfrac{\left(\gamma-1\right)M^2\overline{\dot{Q}}}{2 \gamma \overline{p}\:\overline{u}}\left(2\hat{u}-\left(\gamma-1\right) \hat{p}-\hat{s}\right) + \overline{u}\dfrac{\partial \left(M^2 \hat{u}\right)}{\partial x} - 2M\dfrac{\text{d}M}{\text{d}x}\overline{u} \hat{u} + \overline{u}\dfrac{\partial \hat{p}}{\partial x}\bigg. \bigg)  = 0.
    \end{aligned}
\end{equation}
Algebraic rearrangement simplifies the above equation to,
\begin{equation}
    \begin{aligned}
        &\dfrac{ \textrm{D} I_A}{ \textrm{D} t } + \dfrac{1}{M^2}\bigg( - \left(\gamma - 1\right)M\dfrac{\text{d}M}{\text{d}x}\dfrac{\overline{u}}{\zeta}\left( \hat{u}M^2 + \hat{p} + \dfrac{\hat{s}}{\gamma-1}\right) + \overline{u}\dfrac{\partial}{\partial x}\left( \hat{u}M^2 + \hat{p} + \dfrac{\hat{s}}{\gamma-1}\right)+ \\& \dfrac{\overline{u}}{\gamma - 1} \dfrac{\partial \hat{s}}{\partial x} +  \dfrac{\left(\gamma-1\right)M^2\overline{\dot{Q}}}{2 \gamma \overline{p}\:\overline{u}}\left(2\hat{u}-\left(\gamma-1\right) \hat{p}-\hat{s}\right)\bigg. \bigg)  = 0.
    \end{aligned}
\end{equation}
Using~\eqref{eq:FlowInvariants} we replace $\left( \hat{u}M^2 + \hat{p} +\dfrac{\hat{s}}{\gamma-1}\right)$ with $\dfrac{\zeta I_B}{\gamma - 1}$ and $\hat{s}$ with $I_C$. This gives,
\begin{equation}
    \begin{aligned}
        &\dfrac{ \textrm{D} I_A}{ \textrm{D} t } + \dfrac{1}{M^2}\left( - M\dfrac{\text{d}M}{\text{d}x}\overline{u} I_B + \overline{u}\dfrac{\partial}{\partial x}\left( \dfrac{\zeta I_B}{\gamma - 1}\right)+  \dfrac{\overline{u}}{\gamma - 1} \dfrac{\partial I_C}{\partial x}\right) +  \dfrac{\left(\gamma-1\right)\overline{\dot{Q}}}{2 \gamma \overline{p}\:\overline{u}}\left(2\hat{u}-\left(\gamma-1\right) \hat{p}-\hat{s}\right)  = 0.
    \end{aligned}
\end{equation}
On further carrying out the differential $\dfrac{\partial }{\partial x}\left(\dfrac{\zeta I_B}{\gamma - 1}\right)$ in the above equation, we obtain,
\begin{equation}\label{eq:Pre_LEE1NIfr}
    \begin{aligned}
        \dfrac{ \textrm{D} I_A}{ \textrm{D} t } + \dfrac{1}{M^2}\dfrac{\overline{u}}{\gamma-1}\left(\zeta\dfrac{\partial I_B}{\partial x} - \dfrac{\partial I_C}{\partial x} \right) + \dfrac{\left(\gamma-1\right)\overline{\dot{Q}}}{2 \zeta \gamma \overline{p}}\left(2\hat{u}-\left(\gamma-1\right) \hat{p}-\hat{s}\right)  = 0.
    \end{aligned}
\end{equation}
\item \textbf{Invariant formulation for normalised stagnation temperature fluctuations, $I_B$}:

On multiplying~\eqref{eq:NormalisedXMomentum} by $M^2$, and using ~\eqref{eq:SimpleMach}, we can write,
\begin{equation}
    \begin{aligned}
        &\dfrac{ \partial}{ \partial t }\left(\hat{u}M^2 + \hat{p} + \dfrac{\hat{s}}{\gamma-1}\right) + \overline{u}\dfrac{ \partial}{ \partial x }\left(\hat{u}M^2 + \hat{p} + \dfrac{\hat{s}}{\gamma-1}\right) -\dfrac{1}{\gamma-1}\dfrac{\text{D}\hat{s}}{\text{D} t} - \dfrac{ \partial \hat{p}}{ \partial t } -2\overline{u}M\dfrac{\text{d}M}{\text{d}x}\hat{u} + \\&\dfrac{\overline{u}}{\zeta}\left(M\dfrac{\textrm{d}M}{\textrm{d}x} + \dfrac{\left(\gamma-1\right)\overline{\dot{Q}}}{2 \gamma \overline{p}\:\overline{u}}\right)\left(2\hat{u}-\left(\gamma-1\right) \hat{p}-\hat{s}\right) = 0.
    \end{aligned}
\end{equation}
Again using~\eqref{eq:FlowInvariants}, we can replace $\left( \hat{u}M^2 + \hat{p} +\dfrac{\hat{s}}{\gamma-1}\right)$ by $\dfrac{\zeta I_B}{\gamma - 1}$. $\dfrac{\text{D}\hat{s}}{\text{D}t}$ and $\dfrac{\partial \hat{p}}{\partial t}$ are also substituted using~\eqref{eq:NormalisedEntropy} and~\eqref{eq:NormalisedMassEnergy}, respectively. This gives,
\begin{equation}
    \begin{aligned}
        &\dfrac{ \partial}{ \partial t }\left( \dfrac{\zeta I_B}{\gamma - 1}\right) + \overline{u}\dfrac{ \partial}{ \partial x }\left(\dfrac{\zeta I_B}{\gamma - 1}\right) + \overbrace{\cancel{\dfrac{\overline{\dot{Q}}}{\gamma \overline{p}}\left(\hat{u}+ \gamma\hat{p}\right)}}^{\dfrac{-1}{\gamma - 1}\dfrac{\text{D}\hat{s}}{\text{D}t}} +  \overbrace{\overline{u}\dfrac{\partial}{\partial x}\left( \hat{p} + \hat{u}\right) + \dfrac{\left(\gamma\cancel{-1}\right)\overline{\dot{Q}}}{\gamma \overline{p}}\left(\hat{u}+ \gamma\hat{p}\right) }^{\dfrac{-\partial \hat{p}}{\partial t}} +\\& 2\overline{u}M\dfrac{\text{d}M}{\text{d}x}\hat{u}\left(\dfrac{1}{\zeta}-1\right)  - \left(\gamma-1\right)\dfrac{\overline{u}}{\zeta}M\dfrac{\textrm{d}M}{\textrm{d}x}\left( \hat{p}+\dfrac{\hat{s}}{\gamma-1}\right) + \dfrac{\left(\gamma-1\right)\overline{\dot{Q}}}{2 \zeta \gamma \overline{p}}\left(2\hat{u}-\left(\gamma-1\right) \hat{p}-\hat{s}\right) = 0.
    \end{aligned}
\end{equation}
Further using~\eqref{eq:FlowInvariants}, we replace $\hat{p} +\hat{u}$ with $ I_A + I_C$. This results in:
\begin{equation}\label{eq:IB_Step4}
    \begin{aligned}
        &\dfrac{\zeta}{\gamma-1}\dfrac{ \text{D}I_B}{\text{D} t } + \cancel{\dfrac{\overline{u} I_B} {\gamma-1}\dfrac{ \partial \zeta}{ \partial x }} +  \overline{u}\dfrac{\partial}{\partial x}\left(I_A + I_C\right) - \cancel{ \left(\gamma-1\right)\dfrac{\overline{u}}{\zeta}M\dfrac{\textrm{d}M}{\textrm{d}x}\overbrace{\left( \hat{u}M^2 + \hat{p}-\dfrac{\hat{s}}{\gamma-1}\right)}^{\zeta I_B/\left(\gamma - 1\right)}} + \\ & \dfrac{\gamma\overline{\dot{Q}}}{\gamma \overline{p}}\left(\hat{u}+ \gamma\hat{p}\right) + \dfrac{\left(\gamma-1\right)\overline{\dot{Q}}}{2 \zeta \gamma \overline{p}}\left(2\hat{u}-\left(\gamma-1\right) \hat{p}-\hat{s}\right)  = 0.
    \end{aligned}
\end{equation}
On multiplying the resultant~\eqref{eq:IB_Step4} with $(\gamma-1)/\zeta$ and rearranging, we get,
\begin{equation}\label{eq:Pre_LEE2NIfr}
\begin{aligned}
        &\dfrac{ \textrm{D} I_B}{ \textrm{D} t } + \dfrac{\left(\gamma-1\right)}{\zeta} {\overline{u}} \dfrac{\partial \left(I_{A} +I_C\right)}{\partial x} \\&+ \dfrac{\left(\gamma-1\right)\overline{\dot{Q}}}{\gamma \overline{p} \zeta} \left(\hat{p}\left(\gamma^2-\dfrac{{\left(\gamma-1\right)}^{2} M^2 }{2\zeta}\right) + \hat{u}
        \left(\gamma + \dfrac{\left(\gamma-1\right)M^2}{\zeta}\right) - \hat{s} \dfrac{\left(\gamma-1\right)M^2}{2\zeta}\right)= 0.
        \end{aligned}
\end{equation}
\item \textbf{Invariant formulation for normalised fluctuating entropy, $I_C$:}

Using $I_C = \hat{s}$, we also have from~\eqref{eq:NormalisedEntropy},
\begin{equation}\label{eq:Pre_LEE3NIfr}
\begin{aligned}
        &\dfrac{ \textrm{D} I_C}{ \textrm{D} t } + \dfrac{\left(\gamma-1\right)\overline{\dot{Q}}}{\gamma \overline{p}}\left(\gamma\hat{p} + \hat{u}\right) = 0.
        \end{aligned}
\end{equation}
\end{itemize}

The above equations~\eqref{eq:Pre_LEE1NIfr}, \eqref{eq:Pre_LEE2NIfr} and~\eqref{eq:Pre_LEE3NIfr} can be written in a simplified matrix form as,
\begin{equation}\label{eq:FinalPDE_appendix}
\dfrac{\partial}{\partial t} \mathsfbi{I} \:+\:\overline{u} \mathsfbi{E}_{x} \frac{\partial}{\partial x} \mathsfbi{I}\:+\: \dfrac{\left(\gamma -1\right) \overline{\dot{Q}}}{2\zeta \gamma \overline{p}}\mathsfbi{E}_{s,\:primitive} \mathsfbi{P} = 0,\quad \text{where},
\end{equation}
\begin{equation}\label{eq:Matrix_finalPDE_appendix}
\mathsfbi{E}_{s,\:primitive}=\left[\begin{array}{ccc}
{-\left(\gamma-1\right)}& 2 & -1 \\
2\gamma^2-\dfrac{{\left(\gamma-1\right)}^{2} M^2 }{\zeta} & 2\gamma + \dfrac{2\left(\gamma-1\right)M^2}{\zeta} & -\dfrac{\left(\gamma-1\right)M^2}{\zeta}\\
2\zeta \gamma & 2\zeta  & 0
\end{array}\right],
\end{equation}
and $\mathsfbi{E}_x$ is given in~\eqref{eq:FinalPDE}.

Using~\eqref{eq:DIPMatrix}, we can write, 
\begin{equation}
\mathsfbi{E}_{s,\:primitive} \mathsfbi{P} \equiv \mathsfbi{E}_{s,\:primitive} \left[\boldsymbol{D}_{I}^{P}\right]^{-1} \mathsfbi{I} = \mathsfbi{E}_s \mathsfbi{I},
\end{equation}
where $\mathsfbi{E}_s$ is as given by~\eqref{eq:Matrix_finalPDE}. Thus the final system of linearised Euler equations in terms of flow invariants takes the form as in~\eqref{eq:FinalPDE}.
\newpage
\bibliographystyle{jfm}
\bibliography{jfm}
\end{document}